\documentclass[]{emulateapj}

\usepackage{physics}

\usepackage{amsmath}
\usepackage{natbib}

\newcommand{\msun}{\ensuremath{M_\odot}}

\newcommand{\atf}{AT2017gfo}

\newcommand{\beq}{\begin{equation}}
\newcommand{\eeq}{\end{equation}}

\newcommand{\tday}{\ensuremath{t_{\rm d}}}

\newcommand{\lamc}{\ensuremath{\lambda_{\chi}}}
\newcommand{\ti}{\ensuremath{t_{\rm e}}}
\newcommand{\tio}{\ensuremath{t_{\rm e,0}}}

\newcommand{\Expi}{\ensuremath{\mathcal{E}_i}}

\newcommand{\Am}{\ensuremath{\expval{A}}}

\newcommand{\Zm}{\ensuremath{\expval{Z}}}

\newcommand{\e}{\ensuremath{\epsilon}}
\newcommand{\Eday}{\ensuremath{E_{\rm day}}}
\newcommand{\Et}{\ensuremath{E_{\tau}}}

\newcommand{\Lbol}{\ensuremath{L_{\rm bol}}}

\newcommand{\Mej}{\ensuremath{M_{\rm ej}}}

\newcommand{\tf}{\ensuremath{\tau_{\rm 1}}}
\newcommand{\td}{\ensuremath{\tau}}

\newcommand{\vmax}{\ensuremath{v_{\rm max}}}

\newcommand{\thf}{\ensuremath{t_{\rm r}}}
\newcommand{\qnuc}{\ensuremath{\dot{Q}_{\beta}}}
\newcommand{\Qe}{\ensuremath{\dot{Q}_{e}}}

\newcommand{\qdep}{\ensuremath{\dot{q}_{\rm dep}}}
\newcommand{\qa}{\ensuremath{\dot{q}_{\rm a}}}

\renewcommand{\dd}[2]{\frac{d #1}{d #2}}

\begin{document}

\slugcomment{Draft \today}

\title {Radioactive Heating and Late Time Kilonova Light Curves} 
\shorttitle{Late Time Kilonova Light Curves}
\shortauthors{Kasen and Barnes}

\author{Daniel Kasen} \affil{Departments of Physics and Astronomy, University of California Berkeley\\and Lawrence Berkeley National Laboratory} \email{kasen@berkeley.edu}
\author{Jennifer Barnes} \affil{Columbia Astrophysics Laboratory, Columbia University, New York, NY 10032} \affil{NASA Einstein Fellow}
\email{jlb2331@columbia.edu}

\begin{abstract}
Compact object mergers can produce a thermal  electromagnetic counterpart (a ``kilonova") powered by
the  decay of freshly synthesized radioactive  isotopes.
The luminosity  of kilonova light curves depends  on the efficiency with which beta-decay electrons are thermalized in the ejecta.  Here we derive a simple analytic solution for  thermalization by calculating how electrons accumulate in the ejecta and lose energy adiabatically and via plasma losses.
We find that the time-dependent thermalization efficiency is well described by $f(t) \approx (1 + t/\ti)^{-n}$ where $n \approx 1$ and the  timescale $\ti$ is a  function of the ejecta mass and velocity. For a  statistical distribution of r-process isotopes with radioactive power $\qnuc \propto t^{-4/3}$, the late time  kilonova luminosity asymptotes to $L = f(t) \qnuc \propto t^{-7/3}$ and depends super-linearly on the ejecta mass, $L \propto M^{5/3}$. If a kilonova is instead powered by a single dominate isotope,  we show that the late time luminosity can deviate substantially from the underlying exponential   decay  and eventually  become  brighter than the instantaneous radioactivity due to the accumulation of trapped electrons. Applied to the kilonova associated with the gravitational wave source GW170817, these results imply that a possible  steepening of the observed light curve  at $\gtrsim 7$ days is unrelated to thermalization effects and instead could mark the onset of translucency in a high opacity component of ejecta. 
The  analytic results should be convenient for estimating the  properties of observed kilonovae and  assessing the potential late time detectability of future events.
\end{abstract}

\section{Introduction}

The violent merger of two neutron stars (or a neutron star and a black hole) can eject neutron-rich matter  that, upon decompression, will assemble into heavy nuclei via rapid neutron capture (the \emph{r}-process) \citep{ls76, Eichler+89,Meyer89,Rosswog_99,Freiburghaus+99}. The subsequent radioactive decay of these freshly made  nuclei was predicted to power a thermal electromagnetic transient known as a kilonova \citep{Li&Paczynski98,Metzger+10,Roberts+11, bk13}. Electromagnetic follow-up  of the gravitational wave source GW170817 \citep{Abbott_2017a} appears to confirm the existence of an optical/infrared kilonova with properties in general agreement with theoretical expectations for a neutron star merger \citep[e.g.,][]{Abott_2017b,Arcavi+2017,Chornock+2017,Coulter_2017,Cowperthwaite_17,Drout_2017,Kasen+17,Kasliwal+17,Kilpatrick+17,McCully+17,Nicholl+17,Shappee+17,Smartt+17,Soares-Santos_17,Tanaka+17,Tanvir+17}


Interpreting kilonova observations requires understanding the processes by which radioactive decay particles deposit energy (i.e., ``thermalize") in the ejected material.  Radioactivity produces energetic particles (photons, electrons, alphas and fission fragments) which are only partially absorbed and reradiated as thermal light. The thermalization efficiency declines with time as the ejecta expand and dilute, which substantially influences the evolution of the kilonova light curve.

At early times, the luminosity of kilonovae is  complicated by radiation transport effects related to the diffusion of thermal optical/infrared photons through the opaque ejecta. However, at later times ($\gtrsim $~days to weeks),  the ejecta  become optically thin and the bolometric light curve directly tracks the instantaneous deposition of radioactive energy.   This makes the late time light curves of kilonovae particularly sensitive probes of merger ejecta.  A simple theoretical description of thermalization and emission at these phases would  be useful for estimating the physical properties and detectability of kilonovae.

\citet{Metzger+10} made initial analytic estimates of the thermalization in kilonovae, while \citet{Hoto+16} studied the absorption and potential detectability of r-process gamma rays.  \citet{Barnes+16}  carried out detailed numerical calculations of thermalization efficiency for all \emph{r}-process decay products, including electrons, alpha particles and fission fragments. \citet{Waxman+17} applied an analytic treatment of electron thermalization to model the kilonova that accompanied GW170817.
The steep decline of the  efficiency adopted by \citet{Waxman+17} is in tension with the more gradual decrease seen in the numerical results of  \citet{Barnes+16}, motivating a  deeper analytic description of thermalization.

Here we derive analytic expressions for radioactive heating in kilonova  that account for the several important physical processes at play. 
In particular, charged particles from decay are likely trapped by magnetic fields and  accumulate locally
until they are thermalized. The kilonova luminosity is then not simply a function of the instantaneous decay rate, but rather depends on the accumulated store of  electrons emitted from prior epochs.  We determine this cumulative heating by calculating how electrons  deposit energy in the plasma while simultaneously losing energy due to adiabatic expansion. Because plasma losses roughly follow the  Bethe formula ($dE/dt \propto {E}^{-1/2}\ln E$), electrons deposit energy more effectively as they adiabatically degrade to lower energy $E$. We account for this energy dependence, along with the fact that in beta decay the longer lived nuclei on average emit lower energy electrons.

The above physical processes  were  included in the detailed numerical thermalization calculations of \cite{Barnes+16}. Here we show that, despite the apparent physical complexity, the essential behavior of radioactive heating can be well described by simple and intuitive analytic formulae.
After a description of the decay and thermalization processes in kilonovae (\S\ref{sec:heat}), we derive  solutions for the energy evolution and heating efficiency of suprathermal electrons in an expanding plasma (\S \ref{sec:solution}). 
The analytic results are then generalized to varying radioactivity decay parameters (\S \ref{sec:general}) including  heating
 dominated by a single isotope (\S\ref{sec:single}). 
We provide convenient expressions for the thermalization timescale (\S\ref{sec:time}) and the total (gamma-ray 
plus electron) thermalization efficiency of beta decay (\S \ref{sec:total}). 
In \S \ref{sec:conc} we summarize the most  useful results, which are readily applicable to kilonova modeling, and discuss implications for the kilonova associated with GW170817.

\section{Radioactive Heating in Kilonovae} 
\label{sec:heat}

The material ejected in compact object mergers is expected to consist of heavy  neutron-rich isotopes which primarily undergo beta decay.
If trans-lead nuclei are present, alpha decay and fission may also contribute  to the radioactivity.
Detailed nuclear network calculations have shown that  the radioactive power of r-process material is
approximately described by  a power law \citep[e.g][]{Metzger+10, Roberts+11, Lippuner:2015gwa, Rosswog+17}
\beq
\qnuc(t) \approx 10^{10}~ \tday^{-1.3}~{\rm erg~s^{-1}~g^{-1}},
\label{eq:qheat_num}
\eeq
where $\tday$ is the time since merger measured in days. 

The power-law dependence  of $\qnuc(t)$ has been explained as follows.  The r-process synthesizes a 
multitude of isotopes with a wide range of half-lives. Assuming that the decay times, $\thf$,
of isotopes are roughly equally distributed in log time  \citep{Li&Paczynski98} between $t_{\rm min} < \thf < t_{\rm max}$, the integrated number of decays per unit time is 
\begin{equation}
\dot{N}(t) \approx  \frac{N}{\lambda_r}  \int_{t_{\rm min}}^{t_{\rm max}}  \frac{e^{-t/\thf}}{\thf}    d( \ln \thf) 
\approx 
\frac{N}{\lambda_r}  \frac{e^{-t/t_{\rm max}}}{t}
\end{equation}
where $N$ is the total number of isotopes, $\lambda_r = \ln (t_{\rm max}/t_{\rm min})$ is a normalization factor of the distribution,
and we assumed $t \gg t_{\rm min}$.  For times, $t_{\rm min} \ll t \ll t_{\rm max}$ the  number of decays per unit time per gram is
\beq
\dot{n}(t) = \frac{ \dot{N}(t)}{N \Am m_p} = \left[ \frac{1}{\Am m_p \lambda_r} \right] t^{-1},
\label{eq:Ndot}
\eeq
where $\Am$ is the mean atomic weight of isotopes and $m_p$  the proton mass. 

 The radioactive energy generation rate $\qnuc(t)$  declines more rapidly than  $\dot{n}(t) \propto t^{-1}$ because longer lived isotopes typically have a
lower energy release \citep{Colgate_White_66,Metzger+10,Hoto+16}. From Fermi's theory of beta decay,  the average  energy released in a decay 
approximately follows $E_\beta \propto \thf^{-a}$ where
 $a = 1/5$ in the relativistic beta decay regime.
For the epochs of interest to kilonovae ($\sim$ days), \cite{Hoto+16} show that the non-relativistic or non-relativistic Coloumb regime applies, for which $a = 1/4$ and $a=1/3$ respectively. 
Assuming that isotopes with half-lifes $\thf \approx t$ dominate at time $t$, the energy generation rate per  gram is $\qnuc(t) = \dot{n}(t)  E_\beta(t)$ or 
 \beq
\qnuc(t)  \approx  10^{10} \frac{E_{\beta,d}}{m_e c^2}  \frac{200}{\Am}
\tday^{-(1+a)}~{\rm erg~s^{-1}~g^{-1}},
\label{eq:qheat_an}
\eeq%
where $E_{\beta,d}$ is the average 
energy of a beta-decay at 1~day. The analytic estimate  resembles the numerical result Eq.~\ref{eq:qheat_num} with $a \approx 1/3$.

Beta-decays produce gamma-rays, electrons and neutrinos, only a fraction of which will be absorbed and reradiated as kilonova light.
The neutrinos escape straightaway, while gamma-rays will only be effectively absorbed at early times (see \S\ref{sec:total}). After a few days, the kilonova emission is powered mainly by electrons
depositing energy through impact ionization and excitation of ambient atoms \citep{Barnes+16}. The ionization energy loss rate for non-thermal electrons (ignoring relativistic corrections) is given by the Bethe formula
\beq
\dd{E_{\rm ion}}{t} = - \frac{ \pi q_e^4}{m_e v_e} n_{\rm b}  \ln\left(\frac{E}{\chi} \right), 
\label{eq:ion}
\eeq
 where $\chi$ is the effective ionization potential, $n_b$ the number density of bound electrons, and
$q_e, v_e, E$  are the electron charge, velocity and energy, respectively.  For non-relativistic electrons, the loss rate scales as
$E^{-1/2} \ln(E)$, i.e., lower-energy electrons thermalize more readily. Plasma loss due to  interactions
with free electrons has a similar functional form but is expected to be subdominant
given the  low-ionization state of kilonova ejecta.

Beta-decay electrons also lose energy  as they do work on the expanding ejecta. For kilonovae, the ejecta velocity structure rapidly becomes homologous (velocity proportional to radius) and  the ejecta volume increases as $V \propto t^{3}$. The energy loss to adiabatic expansion is then
\beq
\dd{E_{\rm ad}}{t} = - x \frac{E}{t}, 
\label{eq:adiabatic}
\eeq
where $x=2$ for non-relativistic  and $x=1$ for relativistic particles.  
For purely adiabatic evolution the electron energy follows $E \propto t^{-x}$. The energy lost to  expansion goes into increasing the ejecta kinetic energy and is not available to power the kilonova luminosity. A complete treatment of the electron heating efficiency must therefore account for both adiabatic and ionization loses. 

The propagation of electrons through the kilonova ejecta is hindered by magnetic fields.  The fields  initially present in the neutron star merger will be   diluted by  ejecta expansion,  but the expected  residual field strength ($B \sim \mu$g) still implies an electron  Larmor radius $\sim 10^{6}$ times smaller than the ejecta size \citep{Barnes+16}. Assuming  magnetic fields are not  ordered on large scales, electrons are effectively trapped at a specific mass coordinate and advected with the fluid flow.


\section{Analytic Expression for Thermalization}
\label{sec:solution}

We now derive analytic formulae for the thermalization efficiency of electrons (or other charged particles) in  a homologously expanding medium. We assume electrons are trapped locally by magnetic fields at a fixed Lagriangian coordinate, where the time-dependent density is
\beq
\rho(t) = \frac{3 \Mej}{4 \pi \vmax^3 t^3} \eta,
\eeq
where $\Mej$ is the ejecta mass, $\vmax$ the maximum ejecta velocity, and $\eta$  a dimensionless parameter that depends on the density structure (for a uniform spherical distribution, $\eta = 1$). The corresponding number
density of bound electrons is $n_b(t) = \rho(t) \Zm/\Am m_p$ where $\Zm$ and $\Am$ are the average nuclear charge and weight, respectively, of isotopes, which are expected to be in a low ionization state.

The total energy loss rate of a non-relativistic electron, including both adiabatic (Eq.~\ref{eq:adiabatic} with $x=2$) and ionization (Eq.~\ref{eq:ion})  losses is
\beq
\dd{E}{t} = - 2 \frac{E}{t} - \frac{ \pi q_e^4}{m_e v_e} \frac{ 3  \Mej \eta}{ 4 \pi \vmax^3 t^3} \frac{\Zm}{\Am} \frac{\lamc}{m_p},
\eeq
where $\lambda_\chi = \ln(E/\chi)$. For the moment we take $\lambda_\chi$ to be constant, but in \S\ref{sec:general} will adopt a more general  dependence of the ionization losses. Defining a characteristic thermalization timescale
\beq
\ti =  \frac{1}{\Et^{3/4}} \left[\frac{3}{\sqrt{32} } \frac{ q_e^4 \lamc}{ m_e^{1/2} m_p} \frac{ M \eta}{\vmax^3} \frac{\Zm}{\Am}  \right]^{1/2}
\label{eq:ttherm}
\eeq
we write the energy evolution equation in dimensionless form
\beq
\dd{\e}{\tau} =  - \frac{2 \e}{\tau} - \frac{\e^{-1/2}}{\tau^3},
\label{eq:diffeq}
\eeq 
where $\tau = t/\ti$ and $\e = E/\Et$. Here $\Et$ is the average energy of electrons emitted at scaled time $\tau = 1$.  The value of \ti\ sets the timescale at which  electron thermalization begins to become inefficient; we will give convenient expressions for calculating it in \S\ref{sec:time}. 

Solving the differential equation Eq.~\ref{eq:diffeq} we find the evolution of an electron's energy 
\beq
\e(\td,\tau_0) = \e_0 \left( \frac{\tau_0}{\td}\right)^2 
\left[ 1 - 
\frac{3}{2} \frac{\e_0^{-3/2}}{\tau_0^3} \left( \tau - \tau_0 \right) \right]^{2/3},
\eeq
where $\e_0$ is the initial energy of an electron emitted at time $\tau_0$.
We assume now that the electrons emitted at  $\tau_0$ come primarily
from beta decays with decay times $\thf/t_e \approx \tau_0$.
Following the discussion of beta decay in \S\ref{sec:heat} we write $\e_0 = \tau_0^{-a}$, which gives
\beq
\e(\td,\tau_0) =  \tau_0^{-a} \left( \frac{\tau_0}{\td}\right)^2 
\left[ 1 - 
\frac{3}{2}  \left( \frac{\tau - \tau_0}{\tau_0^{3 - 3a/2}} \right) \right]^{2/3}.
\label{eq:e_eq}
\eeq
For specificity, we adopt $a=1/3$ in what follows but generalize to arbitrary values in \S\ref{sec:general}.

At any given time, the ejecta is heated by the cumulative deposition from  electrons emitted at earlier times.
The oldest electrons still in existence at a time $\tau$ are those emitted at a time $\tf$ such that $e(\td,\tf) = 0$, which is satisfied when
\beq
\tf + \frac{2}{3}  \tf^{5/2} = \td
\eeq
The equation is not readably solvable for \tf\ but the limiting cases can be determined. For $\tf \ll 1$ particles thermalize
 nearly instantaneously,  
$\tf \approx \td$. For $\tf \gg 1$ thermalization is inefficient and 
\beq
 \tf \approx   \left(\frac{3}{2} \td \right)^{2/5}~~{\rm for}~{\tf \gg 1}.
\label{eq:tmin}
\eeq

To derive the instantaneous heating rate per  gram, $\qdep(\td)$, we integrate the plasma losses (Eq.~\ref{eq:ion}) of all existing electrons produced between times $\tf$ and $\td$
\beq
\qdep(\td) =  \Et \int_{\tf}^{\td}  {\dot{n}(\tau_0)} \frac{[\e(\td,\tau_0)]^{-1/2}}{\td^3} d \tau_0.
\label{eq:basic_qdep}
\eeq
The factor of \Et\ is included so that $\qdep(\td)$ has physical units of energy. 
Here $\dot{n}(t)$ is the number of electrons emitted per unit time per  gram which is
taken from  Eq.~\ref{eq:Ndot}, giving
\beq
\qdep(\td) = \frac{\Et}{\Am m_p \lambda_r \ti} \int_{\tf}^{\td}  \frac{[\e(\td,\tau_0)]^{-1/2}}{\tau_0 \td^3} d \tau_0
\label{eq:qdep}
\eeq
and $\e(\td,\tau_0)$ is given by Eq.~\ref{eq:e_eq}. 
The integration must be done numerically in general, but we can determine the behavior in the asymptotic  limit $\tau\gg 1$. Since thermalization is inefficient at these times, the energy of particles degrades primarily adiabatically ($
\e \propto \tau^{-2}$) and we approximate
\beq
\e(\td,\tau_0) \approx   \tau_0^{-1/3} (\tau_0/\td)^2.
\eeq
Integration of Eq.~\ref{eq:qdep} then gives the asymptotic  heating rate
\beq
 \qa(\td) \approx 
\frac{6}{5} 
 \frac{\Et}{\Am m_p \lambda_r \ti}
\frac{1}{\tau^2} 
 \left[ \tf^{-5/6} - \tau^{-5/6} \right].
 \eeq
Working in the limit of weak thermalization, $\tf \ll \tau$, we neglect the second term in brackets  and
 use the limiting value of \tf\  (Eq.~\ref{eq:tmin}) to find
\begin{align}
 \qa(\td) \approx \left( \frac{144}{125} \right)^{1/3}
\left(  \frac{\Et }{\bar{A} m_p \lambda_r \ti} \right) \tau^{-7/3}.
\label{eq:qadep}
   \end{align}
   
 The electron thermalization efficiency is defined as $f(\tau) = \qdep(\tau)/\Qe(\tau)$, where $\Qe(\td)$ is 
 the instantaneous radioactive energy generation rate of electrons (i.e., that fraction of the total beta decay power  $\dot{Q}_\beta$ emitted in the form of electrons)
 \beq
 \Qe(\tau) = \dot{n}(\tau) \Et \tau^{-1/3} = 
 \frac{\Et}{\Am m_p \lambda_r \ti}  \tau^{-4/3}.
 \label{eq:qraw}
 \eeq
Dividing Eq.~\ref{eq:qadep} by Eq.~\ref{eq:qraw} we find the asymptotic thermalization efficiency 
\beq
f_{\rm a}(\td) 
\approx  \left( \frac{144}{125} \right)^{1/3}  \tau^{-1} .
\label{eq:fasy}
\eeq
The coefficient  is close to unity,  so we arrive at the simple result  $f_{\rm a}(\tau) \approx \tau^{-1}$.  

The analytic solution  Eq.~\ref{eq:fasy} applies only at late times ($\tau \gg 1$). At early times ($\tau \ll 1$) particles thermalize efficiently and $f(\tau) \rightarrow 1$. An {\it ad hoc} formula that interpolates between the two limits is
\beq
f(\tau) \approx (1 + \tau)^{-1}.
\label{eq:finterp}
\eeq

\begin{figure}[t]
\includegraphics[width=3.6in]{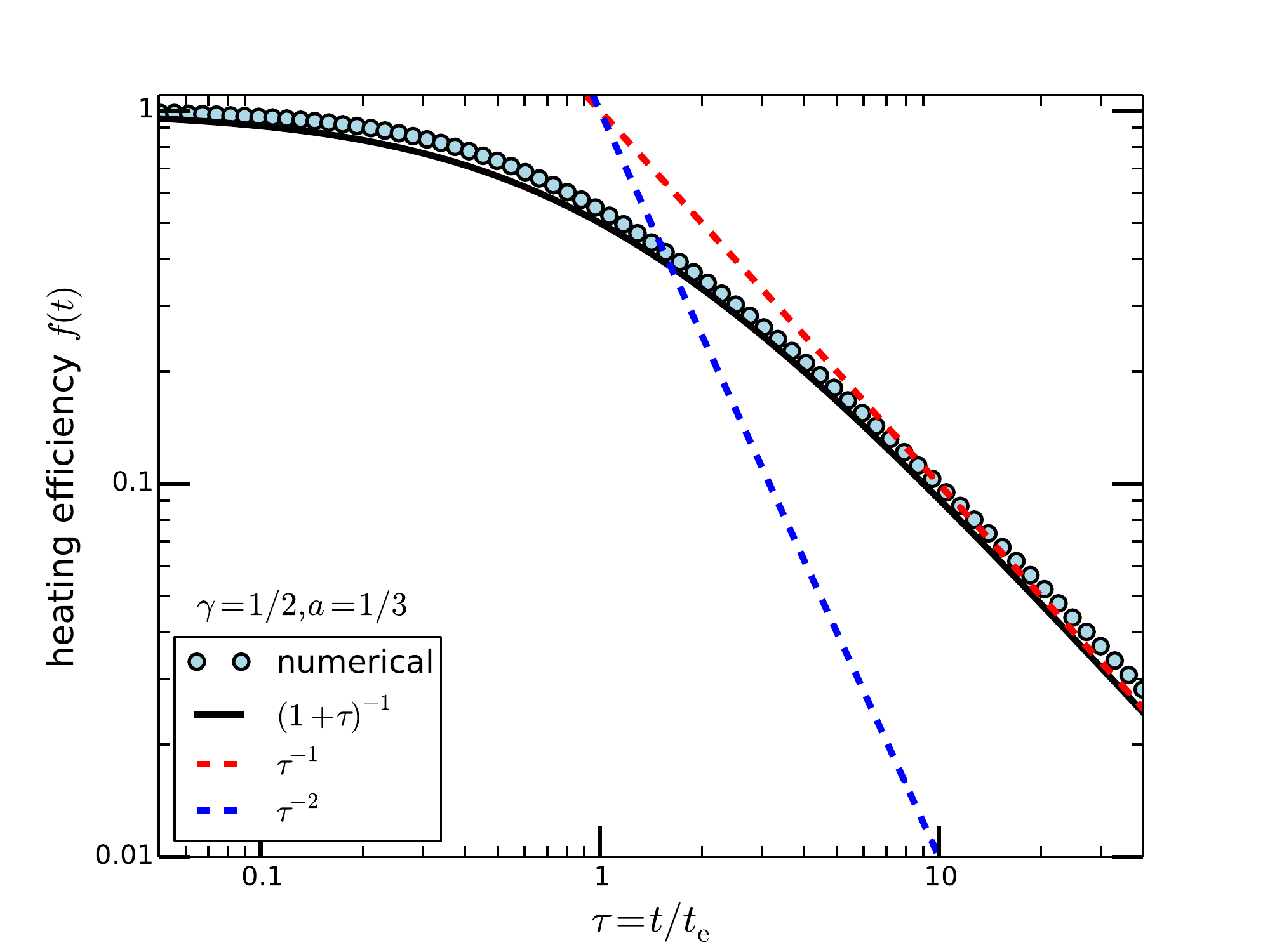}
\caption{Thermalization efficiency of electrons as a function of time for standard parameters.
The numerical result is derived from integrating the plasma loses of accumulated electrons subject to adiabatic
loses. The analytic interpolation formula Eq.~\ref{eq:finterp}
$f(\tau) = ( 1 + \tau)^{-1}$ well approximates the numerical solution, which approaches $f(\tau) \propto \tau^{-1}$ at late times. 
 This calculation uses parameters $\gamma = 1/2, b = 1, a = 1/3, x=2$.
\label{fig:therm}}
\end{figure}

Figure~\ref{fig:therm} shows $f(\tau)$  calculated by numerical integration of Eq.~\ref{eq:qdep} using the  full electron energy
dependence (Eq.~\ref{eq:e_eq}).  The asymptotic behavior approaches the analytic result $f(\tau) \propto \tau^{-1}$. The simple analytic interpolation formula  Eq.~\ref{eq:finterp} reproduces the  numerical solution at all epochs to better than $10\%$. 

The efficiency only gradually approaches the asymptotic behavior $f(\tau) \propto \tau^{-1}$. To quantify the time-dependence
at any instant we can write $f(\tau) \propto \tau^{-n_{\rm eff}(\tau)}$, where 
the effective  exponent  
\beq
n_{\rm eff}(\tau) \approx - \frac{ \partial (\log f)}{\partial (\log \tau)}  \approx \frac{1}{1 + \tau}%
\label{eq:neff}
\eeq
and so  $n_{\rm eff} \le 1$ for all $\tau$. In particular, at the onset of inefficient thermalization ($ \tau = 1$),  the decline rate is only half of the asymptotic result, $n_{\rm eff} = 0.5$. 
This behavior is noticeable in tabulated fits to numerical calculations \citep{Barnes+16},
where  $n_{\rm eff}$ is smaller for models with greater \ti\ (i.e., larger \Mej\ or smaller $\vmax$).

\section{Generalized  Solution} 
\label{sec:general}

The above  thermalization calculation adopted specific dependencies for the electron generation rate, initial electron energies, and 
the plasma loss rate. 
We now derive a more general solution.  
We write the number of electrons generated per gram per unit time as
\beq
{\dot{n}} = B \tau^{-b},
\eeq
where $B$ and $b$ are constants. We take the initial energy of electrons emitted to be $\epsilon(\tau_0) = \tau_0^{-a}$ and  generalize the electron energy equation (Eq.~\ref{eq:diffeq}) to 
\beq
\dd{\e}{\tau} =  - x \frac{ \e}{\tau} - \frac{\e^{-\gamma}}{\tau^3} ,
\label{eq:diffeq_gen}
\eeq 
where $x = 1-2$ quantifies how relativistic the electrons are and $\gamma$ describes the energy dependence of
 loses to the plasma.

In  \S\ref{sec:solution} we adopted default values $a= 1/3, b= 1, x = 2, \gamma = 1/2$.
The actual values likely differ only modestly. 
The  energy dependence of ionization losses may be weaker
than $\gamma = 1/2$
due to the $\lambda_\chi = \log (E_e/\chi)$ term in Eq.~\ref{eq:ion} and 
relativistic corrections.
Inspecting Eq.~\ref{eq:ion} we find that $\gamma \approx  1/4-1/2$ over the energy range
of interest.  

Calculation of the asymptotic thermalization efficiency in the more general formulation can be carried out  in the way described in \S\ref{sec:solution}. We find  
$f_{\rm a}(t) \approx   \tau^{-n}$
where
\beq
n =   1  - a  +   \frac{1 -  \gamma}{1 + \gamma}
- (b-1) \frac{ (2 - a - a \gamma) }
{(x-a) (1 + \gamma) }
\label{eq:general_n}
\eeq
This solution assumes $b + (x-a) \gamma > 1$ and $\gamma > 0$.   
As before, we introduce an {\it ad hoc} interpolation  between the limiting behaviors 
\beq
f(\tau) = (1 + \tau)^{-n}.
\label{eq:localn}
\eeq
For parameters that do not differ much from the defaults, $n$ deviates only modestly from unity. For example, for $b=1, a = 1/3, x= 2$ we find  $n = 1.166$ for $\gamma = 1/3$ 
 and $n = 1.266$ for $\gamma = 1/4$. 

\begin{figure}[t]
\includegraphics[width=3.6in]{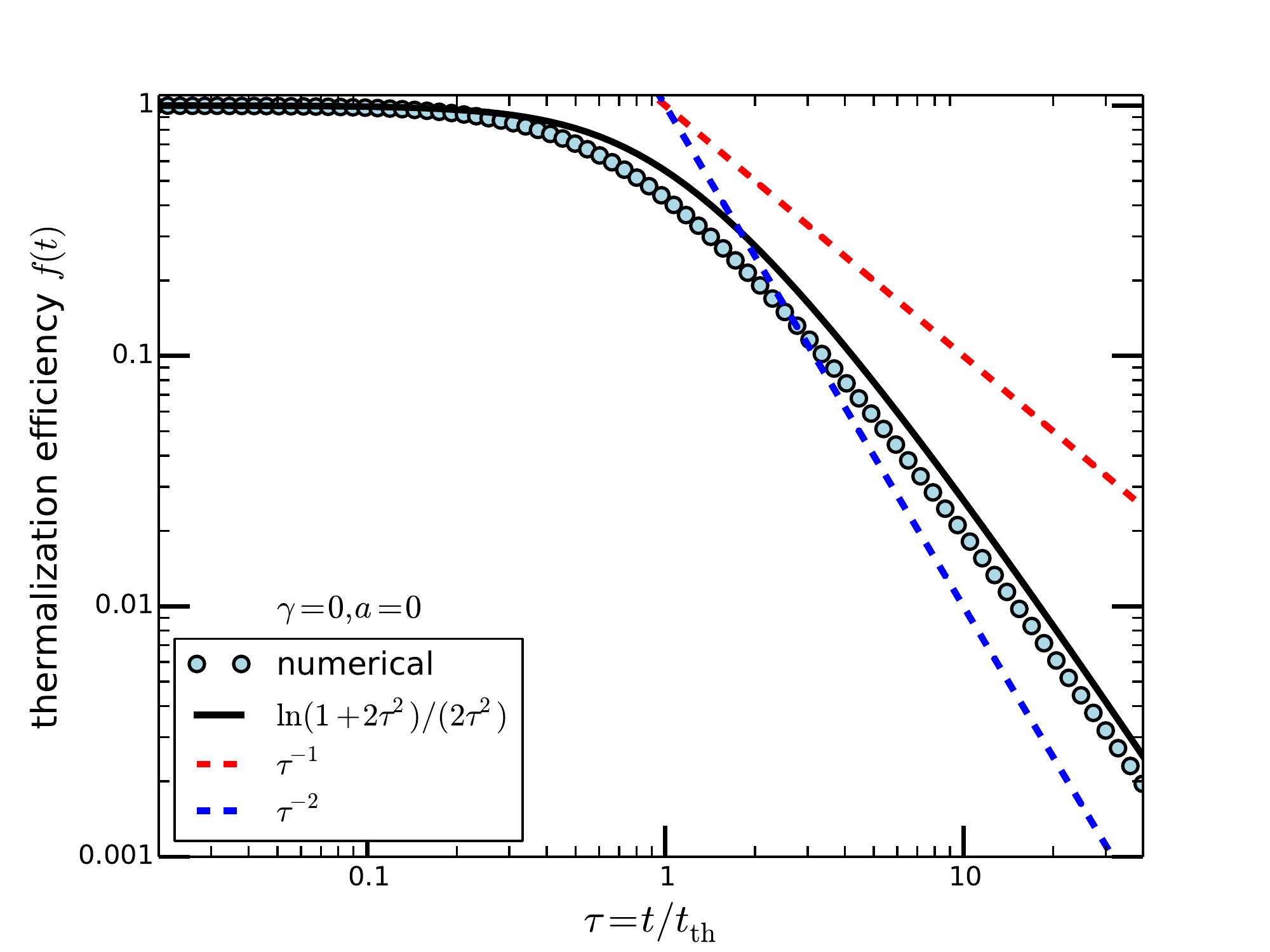}
\caption{
Thermalization efficiency of electrons as a function of time for the case $a = 0, \gamma=0$ (i.e., all electrons emitted with the same energy and plasma loses independent of particle energy). 
The analytic interpolation formula Eq.~\ref{eq:gammazero} reasonably approximates the numerical solution, which declines as $f(\tau) \propto \tau^{-0.8}$ at $\tau \approx 1$ and gradually steepens  to $f(\tau) \propto \tau^{-1.6}$ at $\tau \approx 10$. 
This calculation uses parameters $\gamma = 0, b = 1, a = 0, x=2$.
\label{fig:gammazero}}
\end{figure}
In the limit $\gamma \rightarrow 0$, the approximations applied in the above derivation break down. As an example of the behavior in this regime, we consider the specific case $\gamma=0,~ b=1$ where
integration of
the heating gives 
\beq
\dot{q}_{\rm dep}(\tau) = \frac{\Et}{\Am m_p \lambda_r \ti} \frac{\ln (\tau/\tau_1)
}{\td^{3-a}} .
\eeq
An expression for $\tau_1$ can be determined by solving the energy equation Eq.~\ref{eq:diffeq}
\beq 
\ln \tau =  \ln\tau_1 + \tau_1^2.
\eeq
Calculation of $\tau_1$ must be done numerically, but at late times we have $\tau = \tau_1 e^{\tau_1^2}\gg \tau_1$ and so the asymptotic heating efficiency is
\beq
f_a(\tau) \approx \frac{ \ln(\tau)}{\tau^{2-a}}~~~~({\rm for}~\gamma = 0, b=1).
\eeq
For $a=0$, this efficiency  decays more slowly by a factor $\ln(\tau)$ 
then the $f_a(\tau) \propto \tau^{-2}$ implied by Eq.~\ref{eq:general_n} and adopted by \cite{Waxman+17}.

To describe the full time-dependence of $f(\tau)$ in the limit $\gamma = 0$ we can use an interpolation formula motivated by 
the analytic derivation in \cite{Barnes+16} 
\beq
f(\tau) \approx \frac{\ln(1 +   2 \tau^{2-a})}{2 \tau^{2-a}}~~~~({\rm for}~\gamma = 0, b=1),
\label{eq:gammazero}
\eeq
Figure~\ref{fig:gammazero} shows that Eq.~\ref{eq:gammazero} provides  a reasonable fit to the true numerical solution.
The decay rate at any instant in time can be quantified as $f(\tau) \propto \tau^{-n_{\rm eff}(\tau)}$ with (for $a=0$)
\beq
n_{\rm eff}(\tau) \approx - \frac{ \partial (\log f)}{\partial (\log \tau)} = 2 - \frac{1}{\ln(1 + 2 \tau^2)} \frac{ 4 \tau^2}{1 + 2 \tau^2}
\eeq
which shows that when inefficiency begins to set in ($\tau = 1$)  $n_{\rm eff} \approx 0.8$ which 
steepens to $n_{\rm eff} \approx 1.6$ at very late times $\tau \approx 10$.


\section{Single Isotope Heating}
\label{sec:single}

For some r-process compositions, deviations from a power-law decay rate $\dot{n} \propto t^{-1}$ can occur at times $t > t_{\rm max}$, when the statistical distribution of isotopes cuts off and individual species begin to dominate the radioactive power.
We therefore adapt the previous analysis to derive the heating rate from the exponential decay of a single isotope of decay times \thf. The number of decays per unit time per  gram is now
\beq
\dot{n}_i(t) = \frac{X_i}{A m_p \thf} e^{-t/\thf},
\label{eq:single_n}
\eeq
where $A$ is the atomic mass number and $X_i$  the mass fraction of the isotope. 
The instantaneous radioactive power is  $Q_i(t) = E_i \dot{n}_i(t)$ where 
the energy released per decay $E_i$ is constant with time (i.e., $a=0$). 
The integral for the heating rate (Eq.~\ref{eq:basic_qdep}) becomes for this single isotope case 
\beq
q_i(\td) = 
 \frac{ X_i E_i}{A m_p \thf} \int_{\tau_1}^\tau 
 e^{- \tau_0 \frac{ \ti}{\thf}}
 \frac{  [\e(\tau,\tau_0)]^{-1/2}}{ \tau^3}
     d \tau_0,
     \label{eq:q_ia_integrate}
 \eeq
where we have adopted an  energy loss dependence  $\gamma = 1/2$. The emission time, $\tau_1$, of the oldest living
electrons can be determined from 
electron energy evolution Eq.~\ref{eq:e_eq} with $a=0$ 
\beq
\frac{2}{3} \tau_1^3 + \tau_1 = \tau
\eeq 
and so $\tau_1 = \sqrt[3]{3 \tau/2}$ for  $\tau_1 \gg 1$.

As before, we approximate the late time energy evolution by the adiabatic formula, $\epsilon(\tau, \tau_0) \approx \tau_0^2/\tau^2$, and integrate Eq.~\ref{eq:q_ia_integrate} to find
\beq
q_i(\td) = \frac{X_i  E_i}{A m_p \thf} \frac{1}{\tau^2} \left[ \Expi \left( \tau \ti /\thf \right) - \Expi \left( \tau_1 \ti /\thf \right) \right],
\label{eq:q_i}
\eeq
where \Expi\ is the exponential integral.  In the weak thermalization limit we can neglect the first term in brackets and use the limiting behavior of the exponential integral
$\Expi(x) \approx -e^{-x}/x$, to derive the asymptotic heating rate
\beq
q_{i,a}(\td) =  \frac{ X_i E_i}{A m_p \ti} \left( \frac{2}{3} \right)^{1/3}  
\frac{\exp \left[ - 
\sqrt[3]{ 3 \tau/2} (\ti/\thf)  \right]}{\tau^{7/3}}.
\label{eq:q_ia}
\eeq
The asymptotic thermalization efficiency $f_{i,a}(\tau) = q_{i,a}(\tau)/Q_i(\tau)$ for a single isotope is 

\begin{figure}
\includegraphics[width=3.6in]{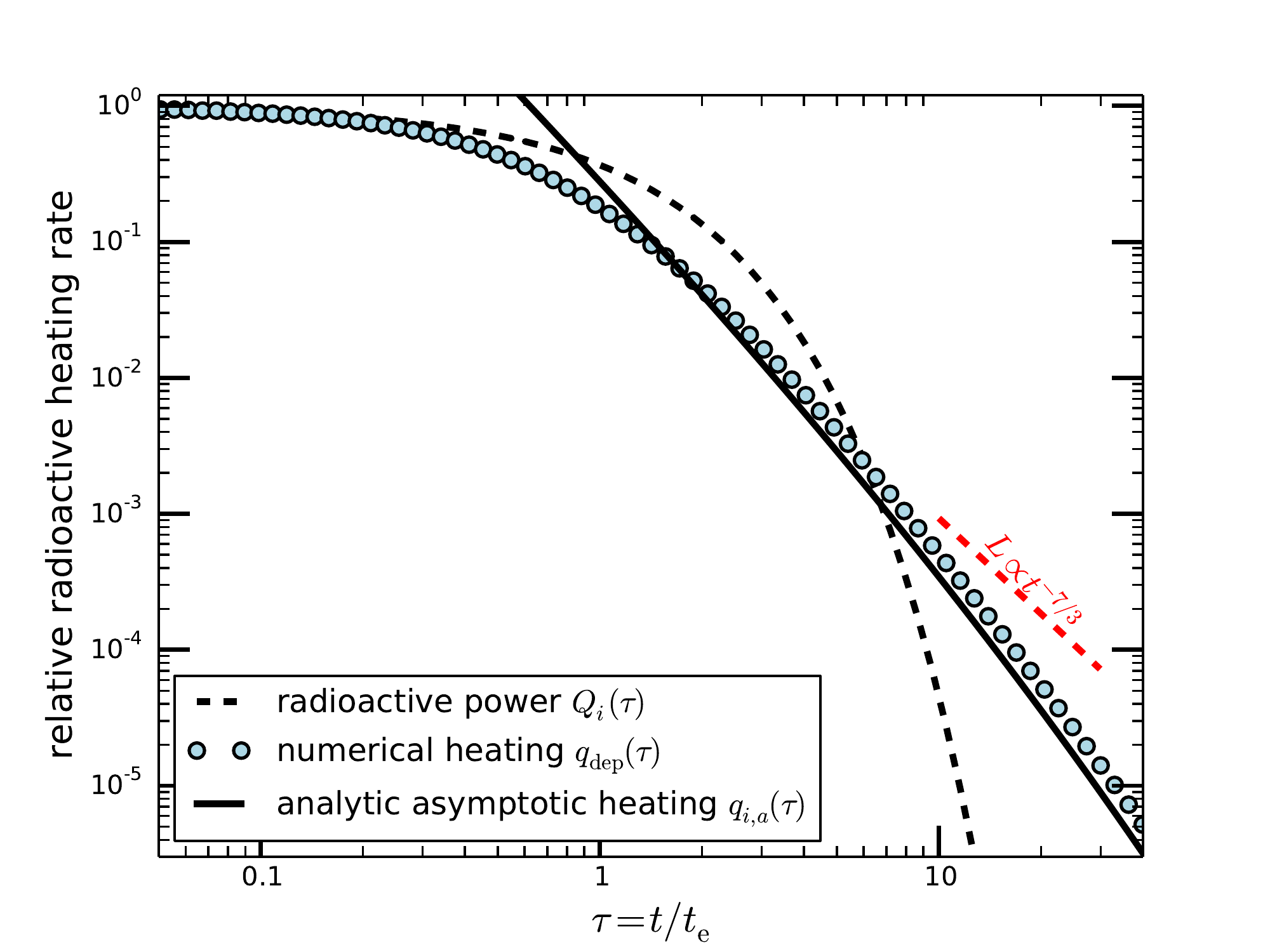}
\caption{Radioactive heating (relative to the value at $t=0$) for a kilonova powered by a single isotope
with a half life taken to be equal to the electron thermalization time \ti. 
The  heating rate $q_i$ (filled circles) deviates substantially from the underlying radioactive decay power $Q_i \propto e^{-t/\thf}$ (dashed black line).
At late times ($\tau \gtrsim 7$) the heating due to electrons accumulated from early epochs {\it exceeds} the instantaneous generation rate $Q_i$, such that the thermalization efficiency is formally greater than one.  
The analytic formula Eq.~\ref{eq:q_ia} (solid black line) reasonably approximates the later time ($\tau \gtrsim 1$) behavior. 
\label{fig:single}}
\end{figure}
\beq
f_{i,a}(\tau) =\frac{\thf}{\ti}
\left( \frac{2}{3} \right)^{1/3}  
\frac{\exp \left[ t_e/\thf (\tau -  
\sqrt[3]{ 3 \tau/2} ) \right]}{\tau^{7/3}},
\eeq
which has the interesting behavior that the  efficiency, at some point, {\it increases} with time 
and eventually will exceed unity. Though perhaps unexpected,  $f(\tau) > 1$ is possible when the heating from  accumulated electrons emitted from previous epochs  dominates over the instantaneous generation rate. This is realized for the steep exponential decay rate of a single isotope, as well as for power-law decay rates when the exponent  $b$ is large enough to give $n < 0$ in Eq.~\ref{eq:general_n}. 

Figure~\ref{fig:single} shows a numerical integration
of the radioactive heating from a single isotope with $\thf = \ti$.
Initially $f(\tau) < 1$, but eventually the integrated heating  due to electrons from earlier epochs  exceeds the
instantaneous radioactive power and   $f(\tau)$ becomes formally greater than one.  The radioactive heating rate  differs substantially from the underlying exponential decay $\propto e^{-t/\thf}$, and is reasonably approximated at times $\tau \gtrsim 1$ by the analytic result Eq.~\ref{eq:q_ia}.
Comparing  this single isotope heating to that of a statistical distribution (Eq.~\ref{eq:qadep}) we see both share a  $\tau^{-7/3}$ dependence, though the single isotope case declines  more steeply 
due to the exponential factor in Eq.~\ref{eq:q_ia}.

\section{Thermalization Timescale}
\label{sec:time}

In our formalism, electron thermalization depends on a single dimensional parameter, \ti, which sets the timescale over which thermalization 
becomes inefficient.  We defined \ti\ in Eq.~\ref{eq:ttherm} as a function of  $\Et$, the energy of electrons emitted at time $t = \ti$. It is convenient to 
rewrite \ti\ in terms of the energy of electrons emitted at some fixed time, say $t = 1$~day
after merger.  Using the time dependence of the electron energy, $E(\tau) = \Et \tau^{-a}$ we rewrite \ti\ from Eq.~\ref{eq:ttherm} as
\beq
\ti = \left[ \frac{\Eday}{m_e c^2} \left( \frac{\ti}{1~{\rm day}} \right)^{-a}  \right]^{-3/4}  \tio
\label{eq:ti_et}
\eeq
where $\Eday$ is the energy of electrons emitted at 1~day and
\beq
\tio =   \left[\frac{3}{\sqrt{32} } \frac{ q_e^4 \lamc}{ m_e^{2} m_p c^3} \frac{M \eta}{\vmax^3} \frac{\bar{Z}}{\bar{A}}  \right]^{1/2}
\eeq
is the thermalization timescale of an electron emitted with  energy $m_e c^2$. 
Solving Eq.~\ref{eq:ti_et} for \ti\ gives the desired expression for \ti
\beq
\ti = \left( \frac{\Eday}{m_e c^2}  \right)^{-3/(4 - 3a) } \left( \frac{\tio}{1~{\rm day}} \right)^{4/(4-3a)}~{\rm days}.
\label{eq:ti_full}
\eeq

To get a sense of the  timescales involved, we scale to values typical for kilonovae. 
For the case $a = 0$ we have
\beq
\ti = 6.8~M_{0.01}^{1/2} v_{0.2}^{-3/2}  \zeta^{1/2}~{\rm days} ~~~~(a = 0),
\eeq 
where $M_{0.01} = \Mej/10^{-2} \msun$ and $v_{0.2} = \vmax/0.2c$.  
For the case $a=1/3$ (our fiducial choice)
\beq
\ti \approx 12.9~
M_{0.01}^{2/3}~ v_{0.2}^{-2}  \zeta^{2/3}
~{\rm days}~~~~~(a = 1/3),
\eeq
above  we have introduced for convenience
the variable
\beq
\zeta =  \eta  \left( \frac{\lambda_{\chi}}{10} \right) \frac{2 \Zm}{\Am}   \left( \frac{\Eday}{m_e c^2}  \right)^{-3/2},
\label{eq:zeta}
\eeq
which is defined such that $\zeta \sim 1$ for typical values.

\section{Total Thermalization Efficiency}
\label{sec:total}

In addition to electrons, beta-decay energy also emerges
as gamma-rays and neutrinos. The neutrinos never thermalize, but gamma-ray deposition can be significant
at early times ($\sim$days).  If a fraction $p_\gamma$ of the energy emerges in gamma-rays and $p_e$ in electrons, the total  
 thermalization efficiency of beta decay is
\beq
f_\beta(t) = p_\gamma f_\gamma(t) +  p_e  f(t),
\eeq
where $f_\gamma(t)$ is the thermalization efficiency of gamma-rays.
Typical fractions for beta decay are $p_e = 0.2, p_\gamma = 0.5$ with the remaining
$p_\nu = 0.3$ emerging as neutrinos \citep[see][]{Barnes+16,Hoto+16}.

Gamma-ray thermalization occurs primarily through inelastic Compton scattering off of bound electrons.  The probability that a gamma-ray emitted at a velocity coordinate $v$ is absorbed in the ejecta is $e^{-\tau(v)}$, where the radial optical depth from $v$ to the surface is, for constant density ejecta
\beq
\tau(v) = \rho \kappa_\gamma (\vmax - v )t,
\eeq
where $\kappa_\gamma$ is the effective absorptive opacity which for $\sim$MeV gamma-rays is approximately $\kappa_\gamma = 0.06~Y_e~{\rm cm^{2}~g^{-1}}$ \citep{Swartz+95}.  The volume averaged optical depth is
\beq
\bar{\tau}_\gamma =   \frac{3}{4 \pi \vmax^3} \int_0^{\vmax} \tau(v) 4 \pi v^2  dv  
= \frac{3 \kappa_\gamma M}{16 \pi \vmax^2 t^2}.
\eeq
Averaging over non-radial gamma-ray trajectories only introduces a small ($\sim 10\%$) correction. 
 
The gamma-ray thermalization efficiency can then be written \citep{Hoto+16, Barnes+16}
\beq
f_\gamma(t) = 1 - \exp \left[ - \frac{t_\gamma^2}{t^2} \right],
\eeq
where $t_\gamma$ is the timescale at which gamma-rays begin to thermalize inefficiently. For constant density ejecta 
\beq
t_\gamma = \left( \frac{ 3 M \kappa_\gamma}{16 \pi \vmax^2} \right)^{1/2}
\approx 0.3~M_{0.01}^{1/2} v_{0.2}^{-1} \kappa_{\gamma,0.02}^{1/2}~{\rm days},
\eeq
where $\kappa_{\gamma,0.02} = \kappa_\gamma/0.02~{\rm cm^{2}~g^{-1}}$.

In outflows with low electron fraction ($Y_e \lesssim 0.15$) the \emph{r}-process can also synthesize significant quantities of translead nuclei \citep[e.g.,][]{Mendoza-Temis+15} and alpha decay will contribute to the radioactive power.   The total heating rate is then
\beq
\dot{q}_{\rm tot}(t) = f_\beta(t) \dot{Q}_\beta(t) + f_\alpha(t) \dot{Q}_\alpha(t),
\eeq
where $\dot{Q}_\alpha$ and $f_\alpha$ are the radioactive power and thermalization efficiency of alpha decay.  For low $Y_e$ outflows, $\dot{Q}_\alpha$ may be from $5\%$ to $40\%$ of $\dot{Q}_\beta$ depending on what nuclear mass model is used.
If many alpha-decaying isotopes are present, the statistical distribution of half-lives should mimic that of the beta-decaying nuclei and decline as a power-law $\dot{Q}_\alpha(t) \propto t^{-1}$. If instead the alpha-decay is dominated by just a few of isotopes, $\dot{Q}_{\alpha}(t)$ will more closely resemble an exponential. 


The analytic formulae for thermalization (\S\ref{sec:general}) can also be applied to alpha decay, for which  $x=2$ and $a=0$. The plasma energy loss rate of alpha decay follows a rough power law with $\gamma = 0.3$ in the energy range of interest \citep{Barnes+16}.  The thermalization efficiency is then described by $f_\alpha(\tau) \approx (1 + t/t_\alpha)^{-n}$  with $n \approx 1.5$, and where the thermalization timescale of alpha decay is roughly $t_\alpha \approx 3 \ti$, due to a higher plasma loss rate. In addition, the alpha decay thermalization efficiency is enhanced relative to beta decays because no alpha-decay energy is lost to neutrinos or weakly thermalizing gamma-rays.

\section{Discussion and Conclusion}
\label{sec:conc}

We have derived simple but effective analytic formulae for calculating the radioactive heating in kilonovae. The  fraction of beta-decay energy  that is absorbed in the ejecta can be estimated using
\beq
f_\beta(t) = p_e \left( 1 + \frac{t}{\ti} \right)^{-n} + p_\gamma \left( 1 - e^{-t_\gamma^2/t^2} \right)
\label{eq:final_f}
\eeq
with $n \approx 1$, and where $p_e \approx 0.2,~ p_\gamma \approx 0.5$ are the fractions of  beta-decay energy emerging in electrons and gamma-rays, respectively. 
The thermalization timescales depend on ejecta mass and velocity as
\begin{align}
\ti &\approx 12.9~M_{0.01}^{2/3}~ v_{0.2}^{-2}~
\zeta^{2/3}~{\rm days} \\
t_\gamma &\approx 0.3~M_{0.01}^{1/2} v_{0.2}^{-1}
~{\rm days}
\label{eq:final_t}
\end{align}
where $M_{0.01} = M_{\rm ej}/0.01~M_\odot$, $v_{0.2} = \vmax/0.2c$ and $\zeta \sim 1$ is given by Eq.~\ref{eq:zeta}.
The summary equations above adopt several default assumptions regarding the radioactive decay behavior; more general results can be found in \S\ref{sec:general}.

Our analytic solutions permit simple estimates of the luminosity of a kilonova at later times. Once the ejecta  have become optically thin to photons, the bolometric luminosity should
track the instantaneous energy deposition rate, $\Lbol(t) \approx \Mej \dot{Q}_\beta(t) f_\beta(t)$, where the radioactive power of a statistical distribution of isotopes is
\beq
\qnuc(t) \approx  10^{10} ~\dot{\epsilon}_{10} \tday^{-\alpha}~{\rm erg~s^{-1}~g^{-1}}
\eeq
where $\dot{\epsilon}_{10}$ is the radioactive energy generation rate at $t = 1$~day in units of $10^{10}~{\rm ergs~s^{-1}~g^{-1}}$.
Nuclear reaction networks find $\alpha \approx 4/3$ and $\dot{\epsilon}_{10} \approx 0.5 - 2.5$, 
with a relatively weak dependence on the ejecta conditions as long as they are sufficiently neutron rich (electron fraction $Y_e \lesssim 0.4$). 
If electrons dominate the heating at these epochs, the predicted bolometric luminosity is (using $p_e = 0.2$)
\beq
\Lbol \approx  4 \times 10^{40} \frac{ \dot{\epsilon}_{10}
  M_{0.01}   t_d^{-\alpha} }{( 1 + 0.08 t_d M_{0.01}^{-2/3} v_{0.2}^2)^{n}}~{\rm erg~s^{-1}}
\label{eq:bolfull}
\eeq
At times late enough that the ejecta are both optically thin and inefficient at thermalizing electrons ($t \gg \ti$)  
  the bolometric luminosity of Eq.~\ref{eq:bolfull} becomes
%
\begin{align}
\Lbol
 \approx 5.2 \times 10^{41}~ \dot{\epsilon}_{10} M_{0.01}^{5/3} v_{0.2}^{-2}  t_d^{-(n + \alpha)}~{\rm erg~s^{-1}}
\label{eq:Lbol}
\end{align}
The late time  luminosity depends super-linearly on \Mej, as a larger ejecta mass produces both greater radioactive power and a higher thermalization efficiency. 
For typical values $\alpha \approx 4/3, n \approx 1$ the  asymptotic dependence is $\Lbol(t) \propto t^{-7/3}$.

 
We further derived analytic heating rates for  radioactivity dominated by a single isotope with an exponential, rather than power-law, time-dependence. This can occur for mildly neutron rich outflows that synthesize only a narrow distribution of isotopes. Interestingly, the late time bolometric luminosity in this case eventually {\it exceeds} the instantaneous radioactive power (i.e., $f(t) > 1$). This is because  the heating from  electrons accumulated from earlier epochs eventually exceeds the generation rate of new electrons. The predicted late time light curves of single isotope kilonovae also have a $\Lbol(t) \propto t^{-7/3}$ dependence (Eq.~\ref{eq:q_ia}) but modulated by an exponential factor that gives a steeper decline.  The non-trivial behavior of $f(t)$ highlights the importance of carefully considering thermalization effects when trying to infer the radioactive source from late time bolometric measurements of kilonovae and supernovae.

We can apply our analytic results to the kilonova \atf\ associated with the neutron star merger GW170817.   The bolometric luminosity at $t = 10$~days was $\Lbol \approx 10^{40}~{\rm erg~s^{-1}}$. 
Taking $\dot{\epsilon}_{10} = 1$, $\alpha=4/3$ and $\vmax = 0.2c$, Equation~\ref{eq:bolfull} gives $\Mej \approx 0.06~M_\odot$, similar to estimates
derived from more detailed modeling of the light curve. Uncertainties in the bolometric correction to the observations, along with the ejecta velocity, density profile, and nuclear heating rate $\dot{\epsilon}_{10}$, however, could introduce errors in \Mej\ at the factor of $\sim 2$ level. 
 
The time-evolution of $f(t)$ is  important for interpreting 
the bolometric  light curve of \atf, which initially declined as  $\Lbol \propto t^{-1}$  then appeared to steepen to $\Lbol \propto t^{-3}$ at times $t \gtrsim 7$~days \citep{Cowperthwaite_17,Drout_2017, Kasliwal+17,Kilpatrick+17,Smartt+17,Waxman+17, Arcavi18,Coughlin+18}.  While this steepening has potentially interesting implications for the kilonova properties,  it may also be an artifact 
of a shifting bolometric correction -- late times observations are available in only a few wavelength bands and
 different published bolometric reconstructions find discrepant  results \citep{Arcavi18}.  

\cite{Waxman+17} ascribe the bolometric steepening in \atf\ to the onset of inefficient  thermalization, 
which they model as a sudden transition from unity to $f(t) = (t/\ti)^{-2}$ for $t > t_e$. Our analysis  indicates that this interpretation is unlikely -- the thermalization efficiency has a weaker asymptotic decline $f(t) = (t/\ti)^{-1}$ and this is only approached gradually. At the onset of inefficiency ($t \approx \ti$, expected to occur $\sim$~weeks after the merger) the dependence is approximately $f(t) \propto t^{-0.5}$ (see Eq.~\ref{eq:neff} and Figure~\ref{fig:therm}) which is too shallow to explain a relatively sharp steepening to $\Lbol \propto t^{-3}$.

A change in the light curve slope could occur at times $t > t_{\rm max}$ when the statistical distribution of isotopes cuts off and one or a small number of decays start  to dominate the underlying radioactive power. The
steeper heating evolution of Eq.~\ref{eq:q_i} then applies.
Nuclear reaction networks for various outflow conditions do show eventual deviation from a power-law  \citep{Rosswog+17}, although this transition
typically occurs at  later times, $t \gtrsim 15$~days.

\begin{figure}
\includegraphics[width=3.6in]{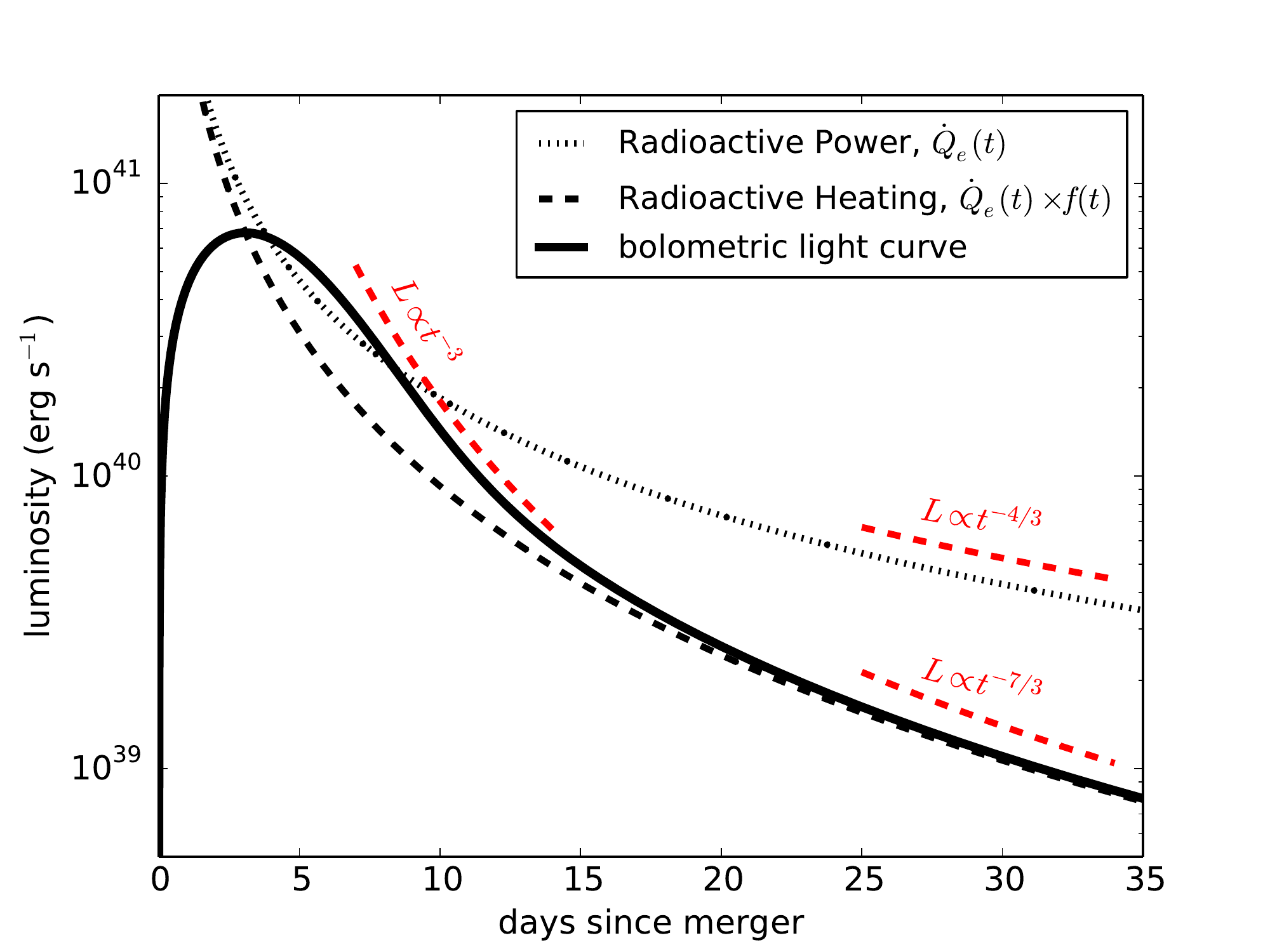}
\caption{Toy analytic light curve of a kilonovae with radioactive power $Q_\beta \propto t^{-4/3}$ and
 a heating efficiency $f(t) = (1 - t/\ti)^{-1}$ with $\ti = 10$~days. The light curve is calculated using a simple Arnett-like one-zone semi-analytic model \citep{Arnett_1982, Kasen&Bildsten10} with an effective diffusion time, $t_d = 5$~days. The relatively steep decline $L \propto t^{-3}$ after peak is due to opacity effects as trapped radiation diffuses out of the ejecta, while the shallower late time decline follows the
asymptotic result $L \propto t^{-7/3}$. 
\label{fig:lc}}
\end{figure}

Another plausible explanation of the light curve steepening of \atf\ is that some significant portion of the   ejecta remained optical  thick to photons for $t \approx 7$ days. As the kilonova ejecta become translucent,  trapped  radiation is released, causing the light curve to decline more steeply then the instantaneous heating rate. This  behavior is familiar from observations of supernova light curves, which show a sharp decline from peak followed by a shallower radioactive ``tail".  We illustrate the effect with a simple analytic model in Figure~\ref{fig:lc}, which captures the bolometric behavior seen in detailed radiation transport calculations \citep[e.g.,][]{Kasen+17, Kilpatrick+17, Tanaka+18, Wollaeger+18}.
For a kilonova to remain  optically thick over $\sim 7$~days requires a high opacity presumably provided by complex lanthanide ions, suggesting that GW170817 synthesized a significant mass of heavy ($A \gtrsim 130$) r-process ejecta \citep{Kasen+13}.  This is consistent with the red colors observed at the later epochs, which are defining signature of lanthanide production \citep{bk13, Tanaka&Hotokezaka13}. 
%

The analytic results derived here provide workable estimates for analyzing and understanding kilonova light curves, but quantitative accuracy requires explicit thermalization transport calculations based on detailed nuclear inputs \cite[e.g.,][]{Barnes+16}.  We have simplified here the complex cascades of beta decay modes which can produce a varying electron spectrum and time-evolution. 
In addition, alpha-decay  is generally  more efficiently thermalized than beta-decay energy and may become significant at late times, in some cases dominating the heating \citep{Barnes+16}.
Quantitative analyses of kilonova observations will require further nuclear experiment and theory to determine the 
detailed nucleosynthesis and decay chains of r-process nuclei.

While the results presented here clarify some aspects of the bolometric emission of kilonovae, the predicted late time  colors and spectra  remain rather uncertain. Once the ejecta have become fully transparent (the ``nebular phase")  deviations from local thermodynamic equilibrium become significant. At these phases, non-thermal beta-decay electrons will play a dominant role in setting the ionization/excitation state of the ejecta.   
The deposited energy may not strictly speaking be ``thermalized"; nevertheless it will presumably be radiated rapidly via some series of optical/infrared atomic transitions.  Although the microscopic processes will be complex in detail,  the simple estimates of the bolometric luminosity presented here are likely to  remain  robust.

\acknowledgments 
We thank E. Waxman, E.~Ofek, M. Coughlin, and A. Jerkstrand for discussions concerning radioactive thermalization. This work was supported in part by the Department of Energy Office of Nuclear Physics grants
DE-SC0018297 and DE-SC0017616, 
and by the Director, Office of Energy Research, Office of High Energy and Nuclear Physics, Divisions of
Nuclear Physics, of the U.S. Department of Energy under Contract No.
DE-AC02-05CH11231. 
JB is supported by the National Aeronautics and Space Administration  (NASA) through the Einstein Fellowship Program, grant number PF7-180162

\bibliographystyle{apj} 

\end{document}